# Phonon Bloch Oscillations

R. Merlin

*The Harrison M. Randall Laboratory of Physics, University of Michigan,
Ann Arbor, Michigan 48109-1040, USA*

The classical motion of a one-dimensional chain of atoms coupled through a specific force function that depends on position shows features very similar to the Wannier-Stark problem of a quantum particle under the combined effects of a periodic lattice potential and a constant electric field. Both problems exhibit localized modes and a ladder of equally-spaced eigenfrequencies, leading to temporal dynamics characterized by periodic oscillations.

The quantum problem of a particle subjected to both a periodic potential and a constant electric field has been the source of much historical controversy [1,2,3]. Following a burst of theoretical activity in the 1970's, it is now fairly well established that, in the absence of interband terms, which induce Zener tunneling [4], the field leads to eigenstate localization and an energy spectrum consisting of a set of equally-spaced levels, known as the Wannier-Stark (WS) ladder. The level separation is given by $eaE_0$, where $e$ is the particle charge, $a$ is the lattice period and $E_0$ is the magnitude of the field. The existence of such ladders is intimately related to the fundamental transport phenomenon of Bloch oscillations [5], that is, dynamics with a period given by $\hbar/eaE_0$, which corresponds to the time it takes for the particle quasi-momentum to move across a full Brillouin zone. Bloch oscillations have been observed in, e. g., semiconductor superlattices [6] and accelerated optical lattices [7]. More recently, it has been shown that these oscillations play a central role in the generation of high harmonics in solids [8]. Here, a relationship is established between the WS problem and the lattice dynamics of a one-dimensional chain of atoms with a stiffness that varies from site to site in a particular form. We find that the classical problem exhibits properties that parallel those of the quantum system, particularly in regard to the existence of a uniform ladder of states and, thus, similar periodic temporal behavior, and the localization of the eigenmodes.

Consider a one-dimensional string of $N$ atoms of two different masses, $m$ and $M$, arranged in succession and let $U_s$ be the displacement of the $s$th atom from its equilibrium position. The substitution $U_s = Q_s e^{-i\omega t}$ in the equations of motion leads to

$$-\omega^2 \mu_s Q_s = -(\alpha_s + \alpha_{s-1})Q_s + \alpha_s Q_{s+1} + \alpha_{s-1} Q_{s-1} \tag{1}$$

where $s = 1, \ldots, N$ is the site index, $\mu_s = M$ ($m$) if $s$ is even (odd) and $\alpha_s$ is the site-dependent force function. In the case where the force is a constant, $\alpha$, the solutions are of the form $Q_s \propto e^{iqs}$ ($q = 2\pi p/N$, with $p = 1,\ldots,N/2$) and the eigenfrequencies are given by

$$\omega_\pm^2 = \alpha \frac{m+M}{mM}\left(1 \pm \sqrt{1 - \frac{4mM}{(m+M)^2}\sin^2 q/2}\right) . \tag{2}$$

The plus and minus signs correspond, respectively, to the optical and acoustic branches.

In the following, we consider a stiffness of the form $\alpha_s = w(s+s_0)^2$ and take $M > m$. Calculations for a wide range of parameter values show that the frequency spectrum divides into two branches, as in the constant force problem. Unlike that case, however, the lower branch exhibits a large fraction of surface states while the upper branch shows a ladder of equally-spaced frequencies and localized eigenmodes [9]. An example of the dispersion and the behavior of the ladder modes are shown, respectively, in Fig. 1 and Fig. 2.

The similarities between these results and those for the quantum problem of a Bloch particle in an electric field are very apparent and quite striking. The numerics indicate that the counterpart to the WS step size, that is, $eaE_0$, does not depend on $N$ or $s_0$ and is approximately given by

$$\Delta\omega \approx \sqrt{4w(2/m + 1/M)} \tag{3}$$

for $M \gtrsim 3m$ [9]. Although not pursued here, it is of interest to note that the dispersion of the upper branch closely follows the functional form of the force in the more general case $\alpha_s \propto (s+s_0)^{2r}$, which gives $\omega_t \propto (2t + s_0 - N)^r$ with $t = N/2, \ldots, N$. Moreover, for sufficiently large $M/m$, this and many other monotonic forms for the stiffness, e. g., $\alpha_s = \beta e^{\kappa s}$, give eigenstates that are localized.

The localization properties of the upper-branch eigenstates can be simply understood by considering the transfer matrix relating a pair of neighboring sites to one of its adjacent pairs. From Eq. (1), assuming that $\alpha_s$ is slowly varying, we obtain

$$\begin{pmatrix} Q_{s+1} \\ Q_{s+2} \end{pmatrix} = \begin{pmatrix} -1 & \eta_s \\ -\eta_{s+1} & \eta_s \eta_{s+1} - 1 \end{pmatrix} \begin{pmatrix} Q_{s-1} \\ Q_s \end{pmatrix} \tag{4}$$

where $\eta_s = 2 - \omega^2 M_s / \alpha_s$. The matrix eigenvalues are $(\eta_{s+1}\eta_s/2 - 1) \pm \sqrt{(\eta_{s+1}\eta_s/2 - 1)^2 - 1}$, and their character determines the behavior of the eigenmodes. A real eigenvalue at a particular site, that is, a site for which $|\eta_{s+1}\eta_s/2 - 1| > 1$, indicates that the mode decays or grows exponentially around the site. If, instead, $|\eta_{s+1}\eta_s/2 - 1| < 1$, the eigenvalue is a complex number of unit modulus and, therefore, one expects the eigenmode to vary weakly from one pair to its neighbors. Taking $\alpha_s \approx \alpha_{s+1}$, these considerations indicate that the solutions to the equations of motion satisfy

$$0 < \left(1 - M\omega^2 / 2\alpha_s\right)\left(1 - m\omega^2 / 2\alpha_s\right) \leq 1 \quad . \tag{5}$$

For a constant force function, this expression gives the allowed frequency regions for the acoustic and optical modes, that is, $\omega^2 < 2\alpha/M$ and $2\alpha/m < \omega^2 < 2\alpha(m+M)/mM$, respectively. If, instead, the force function varies from site to site, Eq. (5) should be interpreted as a relationship between $\alpha_s$ and $\omega^2$, which determines the extent of the eigenmode of that particular frequency. For $\alpha_s = w(s + s_0)^2$, the acoustic-mode condition becomes

$$(s + s_0) > \sqrt{\frac{M}{2w}}\,\omega \tag{6}$$

while the limiting expressions for the optical branch give

$$\sqrt{\frac{mM}{2w(m+M)}}\,\omega < s + s_0 < \sqrt{\frac{m}{2w}}\,\omega \quad . \tag{7}$$

It is thus apparent that the acoustic and optic conditions describe, respectively, extended and localized modes. The calculated eigenmodes in Fig. 2 confirm the existence of bound modes for which the region of localization is precisely determined by Eq. (7).

Returning to Eq. (1), the solutions in the limit $M \to \infty$, are $\omega_j^{(0)} = \sqrt{(\alpha_{2j} + \alpha_{2j+1})/m}$ $\approx (2j+s_0)\sqrt{2w/m}$ [10]. As implicitly mentioned earlier, the numerical calculations for $M \gtrsim 3m$ show that the upper-branch eigenfrequencies are given by $\omega_j \approx \omega_j^{(0)}\sqrt{(1+m/2M)}$ with a very high degree of accuracy (and, we emphasize, regardless of the values of $N$ or $s_0$). To derive an expression for the eigenmodes, we approximate $\alpha_{2j-1}, \alpha_{2j-2} \approx \alpha_{2j}$ in Eq. (1) to obtain the following separate equations for the motions of the two atoms

$$\begin{aligned}(2 - M\omega^2/\alpha_{2j})\Delta_j &\approx \delta_{j+1} + \delta_j \\ (2 - m\omega^2/\alpha_{2j})\delta_j &\approx \Delta_j + \Delta_{j-1}\end{aligned}, \qquad (8)$$

which lead to a simple relationship involving only the displacement of the lighter species

$$\left(2 - m\omega^2/\alpha_{2j} - \frac{2\alpha_{2j}}{2\alpha_{2j} - M\omega^2}\right)\delta_j \approx \frac{\alpha_{2j}}{2\alpha_{2j} - M\omega^2}\left(\delta_{j+1} + \delta_{j-1}\right). \qquad (9)$$

Here $\Delta_j = Q_{2j}$, $\delta_j = Q_{2j-1}$ and $j = 1, \ldots, N/2$ is the cell index. Assuming that $M \gg m$, we have for $\omega^2 = 2\alpha_{2j_0}/m$

$$2(\alpha_{2j} - \alpha_{2j_0})\delta_j \approx -\frac{m\alpha_{2j}^2}{2M\alpha_{2j_0}}(\delta_{j+1} + \delta_{j-1}) \quad . \qquad (10)$$

The previous discussion indicates that the ladder modes are localized in the narrow region defined by Eq. (7). Therefore, we can approximate $\alpha_{2j} \approx \alpha_{2j_0} + 4w(s_0 + 2t)(j - j_0)$ to finally get

$$(j - j_0)\delta_j \approx -\eta(\delta_{j+1} + \delta_{j-1}) \quad , \qquad (11)$$

where $\eta = \dfrac{m}{16M}(2j_0 + s_0)$. This equation has the exact solution

$$\delta_j \approx J_{j-j_0}(-2\eta) \tag{12}$$

which is of the same form as the localized states of the WS ladder [2]. For the heavier atoms, we replace the above expression in Eq. (8) and find

$$\Delta_j \approx \dfrac{m}{2M}\left[J_{j-j_0+1}(-2\eta) + J_{j-j_0}(-2\eta)\right] \quad . \tag{13}$$

As shown in Fig. 3, the analytical expressions give displacements that are in fairly good agreement with the results of numerical simulations. Moreover, the fact that the Bessel functions and, thus, the eigenmodes decay exponentially outside the interval $|j - j_0| \leq \pi\eta$ is entirely consistent with the bounds from Eq. (7).

In summary, we have shown that the properties of a quadratically perturbed diatomic chain resemble those of the quantum WS problem in that they both show localized eigenmodes and a ladder of equally spaced levels. The existence of such a ladder guarantees that the classical motion of the chain must be periodic in time if restricted to an arbitrary superposition of upper-branch modes, much in the same way that the dynamics of a quantum particle in a single band manifests itself in the form of Bloch oscillations. Practical implementations of these ideas are likely to involve, e. g., semiconductor superlattices [11], alloys [12] with a graded composition, optical lattices [13] or acoustic metamaterials [14].

# REFERENCES

[1] G. H. Wannier, in Elements of Solid-State Theory (Cambridge Unv. Press, Cambridge, U. K., 1959), pp. 190-193.

[2] H. Fukuyama, R. A. Barit and H. C. Fogedby, Tightly Bound Electrons in a Uniform Electric Field, Phys. Rev. B **8**, 5579 (1973).

[3] G. C. Stey and G. Gusman, Wannier-Stark ladders and the energy spectrum of an electron in a finite one dimensional crystal, J. Phys. C: Solid State Phys. **6**, 650 (1973).

[4] Y. Gefen, E. Ben-Jacob, and A. O. Caldeira, Zener transitions in dissipative driven systems, Phys. Rev. B **36**, 2770 (1987)

[5] F. Bloch, Über die Quantenmechanik der Elektronen in Kristallgittern, Z. Phys. **52**, 555 (1928).

[6] C. Waschke, H. G. Roskos, R. Schwedler, K. Leo, H. Kurz, and K. Köhler, Coherent [Lett. **70**, 3319 (1993).

[7] M. Ben Dahan, E. Peik, J. Reichel, Y. Castin, and C. Salomon, Bloch Oscillations of Atoms in an Optical Potential, Phys. Rev. Lett. **76**, 4508 (1996); S. R. Wilkinson, C. F. Bharucha, K. W. Madison, Q. Niu, and M. G. Raizen, Observation of Atomic Wannier-Stark Ladders in an Accelerating Optical Potential, Phys. Rev. Lett. **76**, 4512 (1996); B. P. Anderson and M. A. Kasevich, Macroscopic Quantum Interference from Atomic Tunnel Arrays, Science **282**, 1686 (1998)

[8] S. Ghimire, A. D. DiChiara, E. Sistrunk, P. Agostini, L. F. DiMauro, and D. A. Reis, Observation of high-order harmonic generation in a bulk crystal, Nature Phys. **7**, 138 (2011); O. Schubert, M. Hohenleutner, F. Langer, B. Urbanek, C. Lange, U. Huttner, D. Golde, T. Meier, M. Kira, S. W. Koch and R. Huber, Sub-cycle control of terahertz high-harmonic generation by dynamical Bloch oscillations, Nature Photon. **8**, 119 (2014).

[9] When the ratio $N/s_0$ is held constant, the gap between the two branches as well as the fraction of localized states in the upper branch remain unchanged as $N \to \infty$ whereas, at a fixed $s_0$, both variables decrease as $m/M$ or $N$ increases. As the value of $m/M$ exceeds a certain threshold or as $N$ becomes sufficiently large, the gap disappears and some ladder and surface states become nearly degenerate. However, the separation between consecutive eigenfrequencies stays the same.

[10] To get a perfect equally-spaced ladder, one should have used $\alpha_s = w[(s+s_0)^2 - (s+s_0)]$. The additional term introduces very minor changes in the numerics if $s_0 \gg 1$.

[11] B. Jusserand and M. Cardona, Raman spectroscopy of vibrations in superlattices, in Light Scattering in Solids V, ed. by M. Cardona and G. Güntherodt, Topics in Applied Physics **66** (Springer, Berlin, 1988).

[12] H. Harima, Properties of GaN and related compounds studied by means of Raman scattering, J. Phys.: Condens. Matter **14**, R967 (2002).

[13] Y. Guo, R. M. Kroeze, B. P. Marsh, S. Gopalakrishnan, J. Keeling and B. L. Lev, An optical lattice with sound, Nature **599**, 211 (2021).

[14] G. Ma and P. Sheng, Acoustic metamaterials: From local resonances to broad horizons, Sci. Adv. **2**, e150159 (2016).

FIGURE CAPTIONS

FIGURE 1 – Eigenfrequencies as a function of (*i*) the eigenvalue index (black curves) and (*ii*) the frequency difference between consecutive eigenmodes (red curves) for $N = 10^3$, $M = 10$ and $m = 1$; $\alpha_s = w(s + s_0)^2$ with $w = 10^{-4}$ and $s_0 = 2 \times 10^3$. The gray line is $(2t - N + s_0)\sqrt{2w(1/m + 1/2M)}$; $t = 1,...,N$. The vertical dashed line, at the edge of the Brillouin zone of the uniform lattice, separates the acoustic-like from phonon-ladder modes while the inflection point at $t \approx 350$ signals the boundary between extended and surface states of the lower branch.

FIGURE 2 – $|Q_s|^2$ vs. site index (*s*) for phonon-ladder modes of frequency (left to right) 30.18, 31.92, 36.26, 39.89 and 42.06. Curves have been shifted vertically by an amount equal to the corresponding eigenfrequency. Parameters are the same as for Fig. 1. The gray line is $\sqrt{2w(1/m + 1/2M)} \, (s + s_0)$.

FIGURE 3 – Eigenvector displacements for the ladder mode of index $t = 600$ ($\omega = 31.881$), shown separately for the (a) lighter and (b) heavier atoms. Parameters are the same as for Fig. 1. The orange curves are, respectively, Eq. (12) and Eq. (13) with $j_0 = 100$. Circles are from matrix diagonalization calculations. Note the different scales, which correlate with the site mass. To better match the calculations, the analytical curves were shifted 3 unit cells to the right.

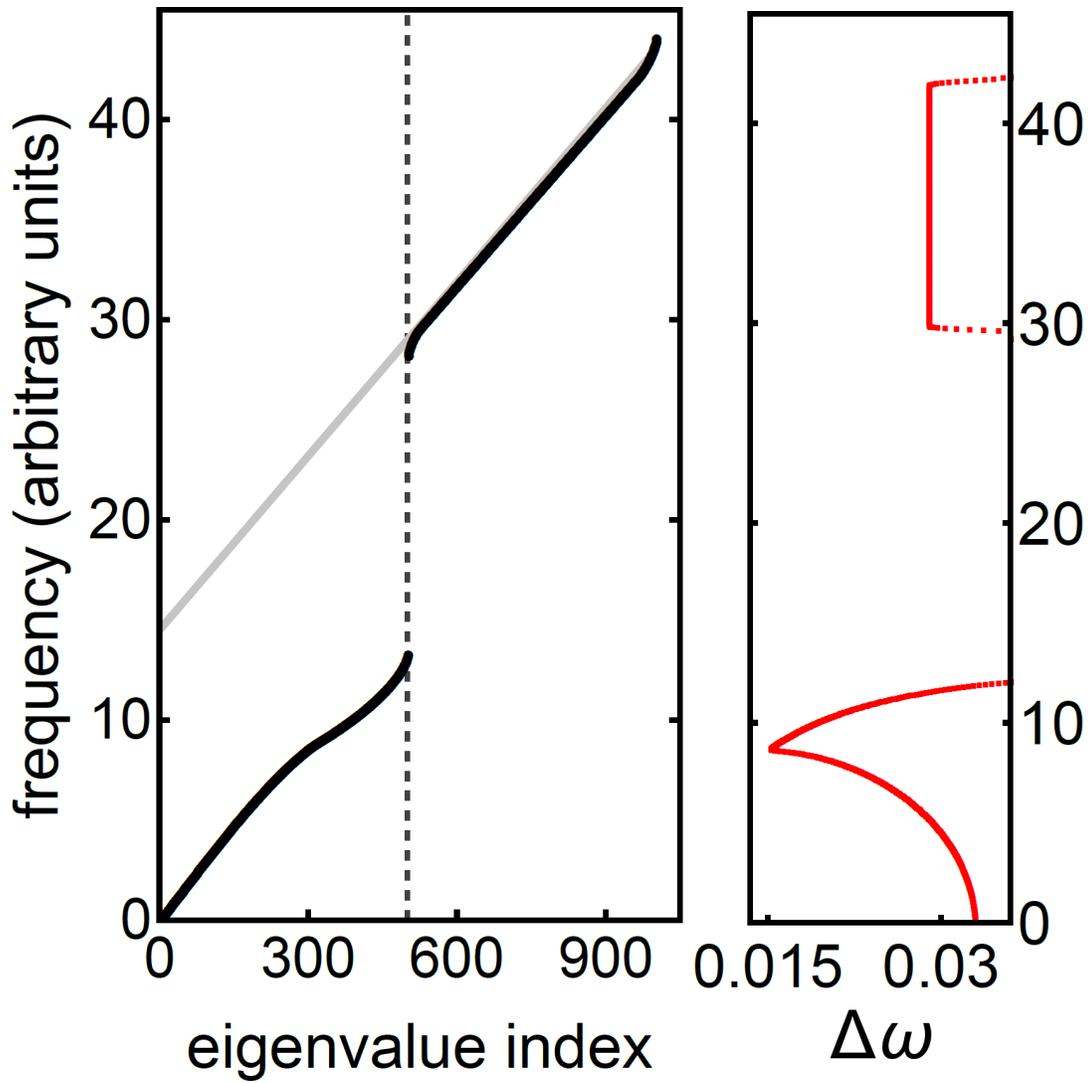

FIGURE 1

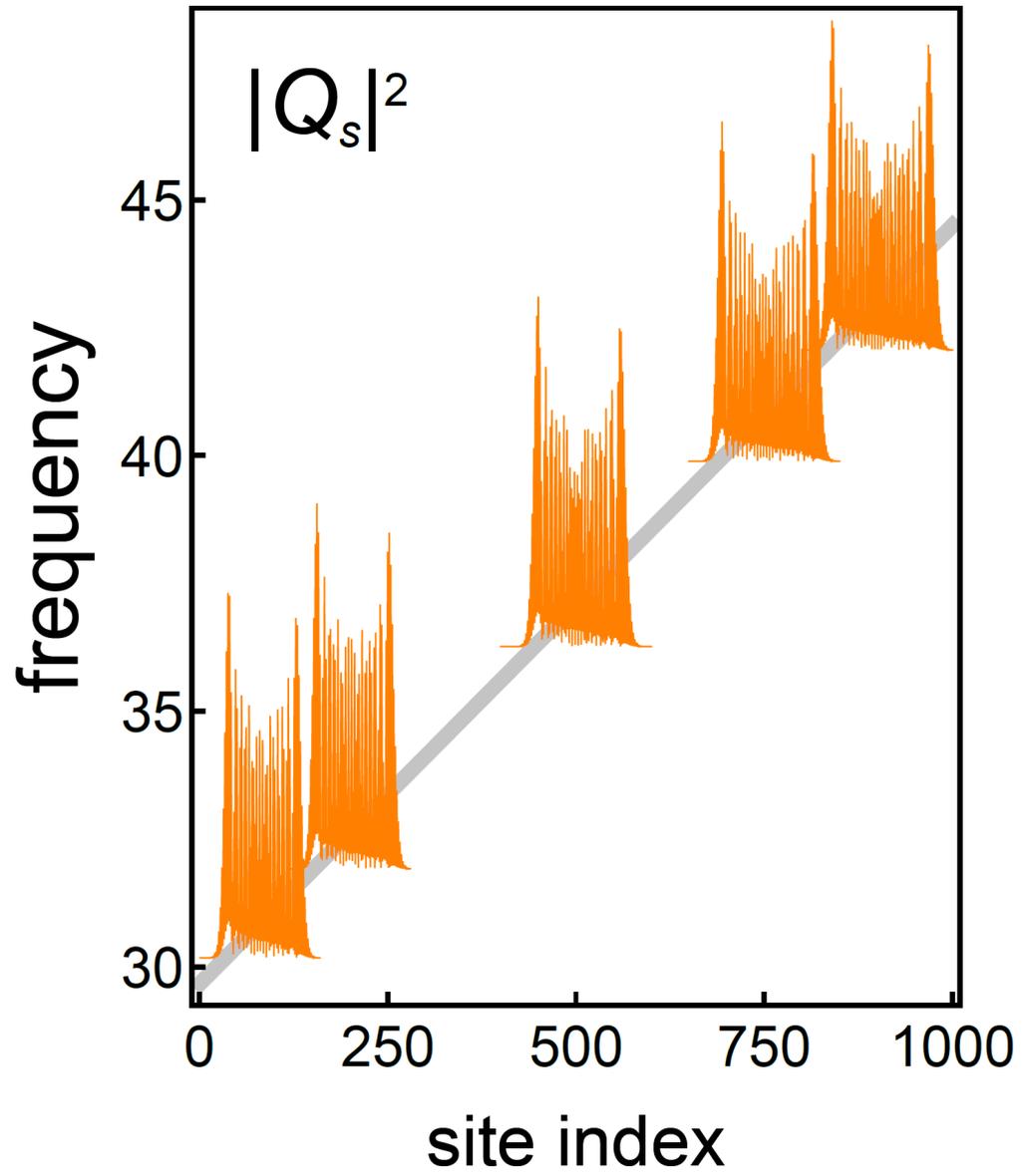

FIGURE 2

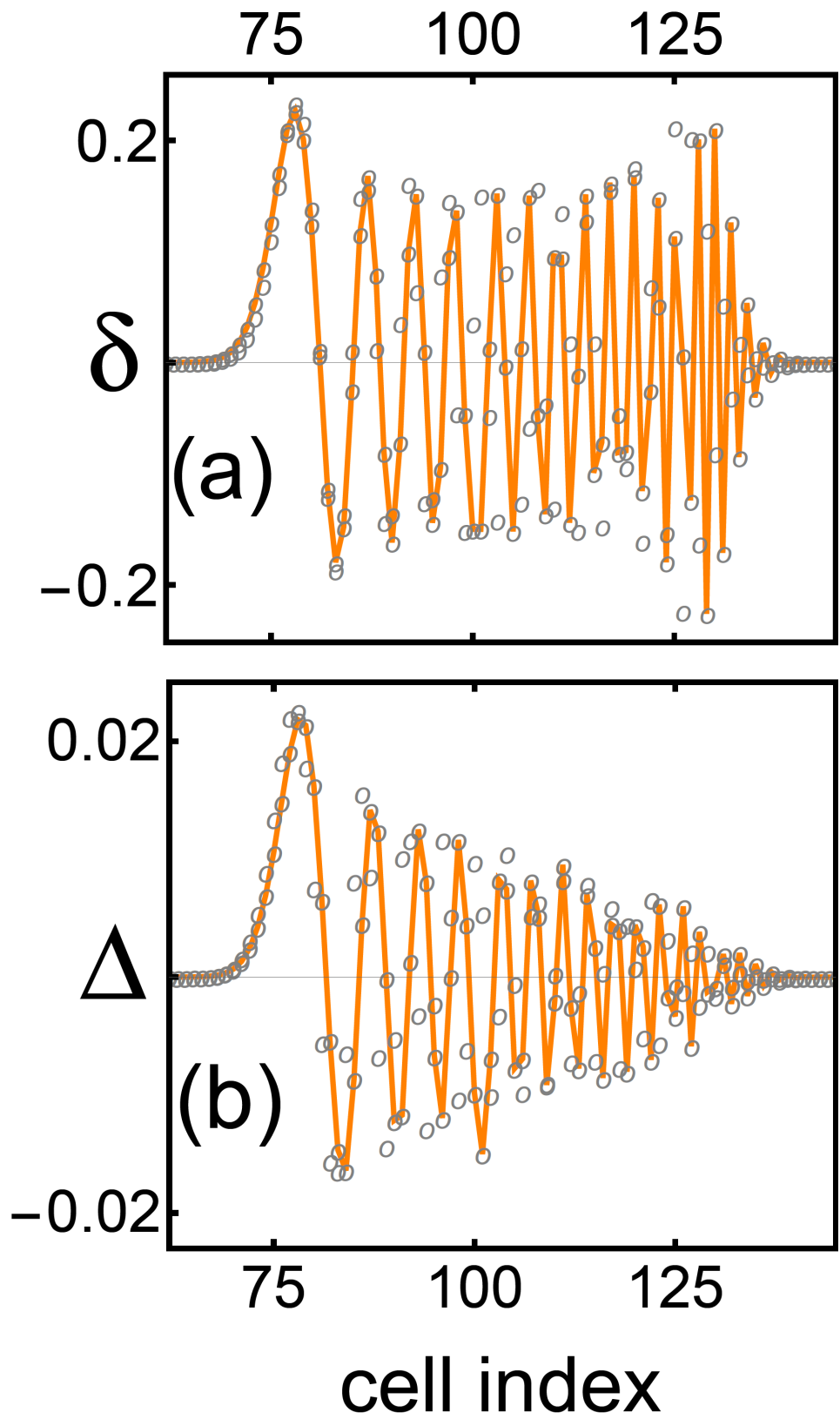

FIGURE 3